\documentstyle[floats, epsfig, prb, aps]{revtex}

\draft

\begin{document}

\twocolumn[\hsize\textwidth\columnwidth\hsize\csname @twocolumnfalse\endcsname

\title{Topological Phase Diagram of a Two-Subband Electron System}

\author{X.~Y.~Lee and H.~W.~Jiang} 
\address{Department of Physics and Astronomy, University of California
  at Los Angeles, Los~Angeles,~CA~90095}

\author{W.~J.~Schaff}
\address{Department of Electrical Engineering, Cornell University,
  Ithaca,~NY~14853} 

\date{\today}

\maketitle

\begin{abstract}
We present a phase diagram for a two-dimensional electron system with
two populated subbands. Using a gated GaAs/AlGaAs single quantum well,
we have mapped out the phases of various quantum Hall states in the
density-magnetic filed plane. The experimental phase diagram shows a
very different topology from the conventional Landau fan diagram. We
find regions of negative differential Hall resistance which are
interpreted as preliminary evidence of the long sought reentrant
quantum Hall transitions. We discuss the origins of the anomalous
topology and the negative differential Hall resistance in terms of the
Landau level and subband mixing. 
\end{abstract}

]

\narrowtext 

Extensive works has been carried out on modulation-doped GaAs/AlGaAs
heterostructures containing a two dimensional electron gas (2DEG)
within the framework of quantum Hall effect.\cite{DasSarma} In most of
these structures, only one subband is populated. Even though studies
of heterostructures with two populated electric subbands have a long
history,\cite{Stormer} the inherent additional inter-subbands
scattering has precluded two-subband system from being a primary
candidate to study various aspects of quantum Hall effect. Most of the
investigations carried out thus far, not surprisingly, have focused on
scattering\cite{scattering} while others dealt with population
effects.\cite{levels} Recently, it has become increasingly apparent
that in one-band systems, disorder induced Landau level mixing can
play a critical role in the evolution of the quantum Hall effect,
especially in the regime of vanishing magnetic fields. Landau level
mixing and its effects have been the subjects of numerous recent
experimental \cite{Glozman1,Glozman2,Kravchenko} and
theoretical \cite{llmixing} studies. Similarly, in a two-subband
system, crossing of Landau levels of the two different subbands can
lead to substantial mixing even in relatively strong magnetic fields.
The consequences of Landau level mixing on the topology of the phase
boundaries between different quantum Hall states \cite{Kivelson} in the
two-band system are expected to be surprising and possibly profound.

To explore some of these consequences, we have conducted a systematic
magneto-transport study on gated, modulation-doped GaAs/AlGaAs single
quantum well samples in which there are two populated subbands. We
have constructed a topological phase diagram of the two-band system.
We found this phase diagram to be considerably more complex than the
conventional Landau fan diagram. One of the spectacular consequences
of its unusual topology is that there are multiple reentrant quantum
Hall transitions. We have observed negative differential Hall
resistance in certain regions of the density-magnetic field plane (the
$n$-$B$ plane).  The negative differential Hall resistance, in our
opinion, is indicative of the reentrant quantum Hall transition.

The sample used in this study is a symmetrical modulation-doped single
quantum well with a width of 250~\AA. Two Si $\delta$-doped layers
($n_{d}= 8 \times 10^{11} \mbox{ cm}^{-2}$) are placed on either side
of the well. There is a 200~\AA~spacer between the $\delta$-doped
layer and the well on each side. Heavy doping creates a very dense
2DEG, resulting in the filling of two subbands in the well. As
determined from the Hall resistance data and Shubnikov-de Hass
oscillations, the total density is $n = 1.21 \times 10^{12}
\mbox{cm}^{-2}$. The higher subband has a density $n_1 = 3.3\times
10^{11} \mbox{ cm}^{-2} $ while the lower subband has a density of
$n_2 =8.8 \times 10^{11} \mbox{ cm}^{-2}$ at $B$ = 0. The electron
mobility at zero gate voltage is about $8~\times
10^4~\text{~cm}^2/\text{V-s}$. The samples are patterned into Hall
bars with a $3:1$ aspect ratio using standard lithography techniques.
An Al gate was evaporated on top so that by applying a negative gate
voltage, the carrier density can be varied continuously. A total of 9
samples with different lengths (varying from 30~$\mu$m to 3~mm) were
studied systematically. For consistency, we present the data from only
one sample here. During the experiment, the sample was thermally
connected to the mixing chamber of a dilution refrigerator. Magnetic
fields up to 12~T were applied normal to the plane containing the
2DEG. Standard lock-in techniques with an excitation frequency of
about 13~Hz and a current of 10~nA were employed to carry out the
magnetoresistance measurements.

A typical trace of the diagonal resistivity $\rho_{xx}$ and the Hall
resistivity $\rho_{xy}$ as a function of $B$ at a temperature of 70~mK
is shown in Fig.~1. The integer numbers in the figure identify each
quantum Hall state by its quantized value in the unit of $h/e^{2}$
(i.e., $S_{xy}= (h/ e^{2})/R_{xy} $). The peaks in $\rho_{xx}$
represent the positions of the delocalized states and together they
mark the phase boundaries between various quantum Hall states. This
criterion was used to construct the experimental phase diagram in Fig.~3.

\begin{figure}[tb]
  \epsfig{file = 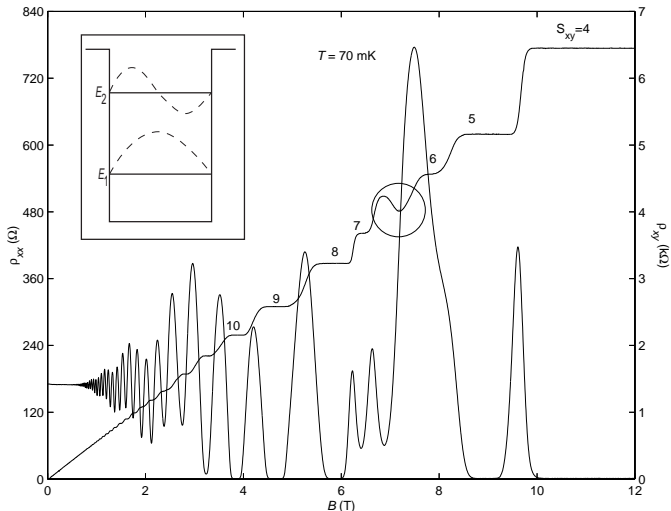, width = 9cm, clip=} 
  \caption{Longitudinal
    and Hall resistivity as a function of the magnetic field.  Trace
    was taken at a fixed gate voltage of -0.41~V at $T$ = 70~mK.
    Inset: schematic band diagram of the quantum well with two
    populated subbands.}  
  \label{trace} 
\end{figure} 

Before presenting the experimental phase diagram, it is useful to
discuss what one should expect for the simplest case in the absence of
Landau level mixing. Using the energy separation between the two
subbands for the present sample, we plot in Fig.~2a the energy $E$ as
function of magnetic field. The corresponding positions of the
delocalized states in the $n$-$B$ plane can be calculated and the
resulting phase diagram is displayed in Fig.~2b. From Fig.~2b, one can
see that the electrons fill the Landau levels of upper and lower
subbands in alternating fashion as the magnetic field is increased. In
this case, the phase diagram has an ordinary ``fan-like'' appearance
identical to that for a single band system.

\begin{figure}[tb]
  \epsfig{file = 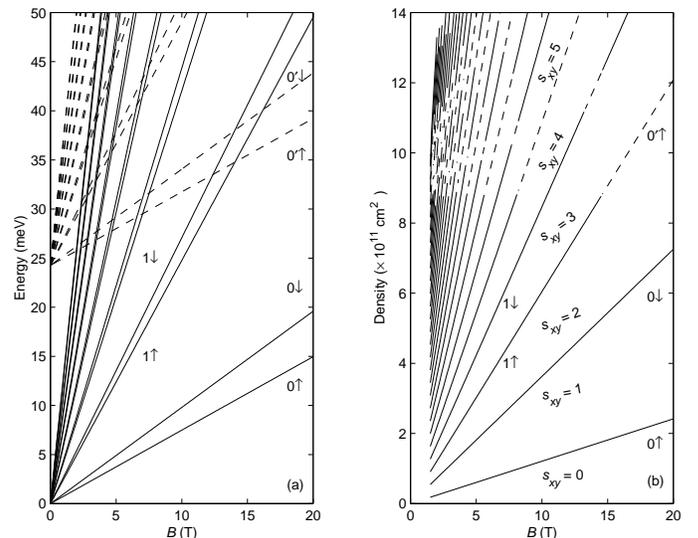, width = 9cm, clip=} 
  \caption{(a) Energy versus magnetic field for the two-subband
    system, and (b) A map of the delocalized states in the density -
    magnetic field plane expected for the disorder free case.  Solid
    lines represent the states populate the first subband and the
    dashed lines are the ones for the second subband.  Numbers
    indicate the association with the different Landau level indices.}
  \label{schematic}
\end{figure} 

The actual phase diagram is, however, very different from this simple
picture. We present in Fig.~3 an experimental phase diagram. The
density on the right axis is related to the gate voltage on the left
axis by a linear relation determined by the sample geometry. To
construct the phase diagram, we have swept both the gate voltage,
i.e., the carrier density, at a fixed magnetic field (a
``$V_g$-scan''), and the field at a fixed gate-voltage (a
``$B$-scan''). Each peak in $\rho_{xx}$ corresponds to a single data
point in the phase diagram. The data points represent the phase
boundaries between various quantum Hall liquid states. Limited by the
base temperature of our cryostat, the plateaus in regions very near to
the intersections of phase boundary lines are normally not well
resolved. In such cases, we follow the evolutionary development of the
plateaus away from these places to assign sensible values of $S_{xy}$.
Moreover, we could not determine the phase boundaries reliably at low
magnetic fields ($B\le ~2~T$) as the peak becomes progressively more
difficult to resolve in a decreasing magnetic field.

\begin{figure}[tb]
  \epsfig{file = 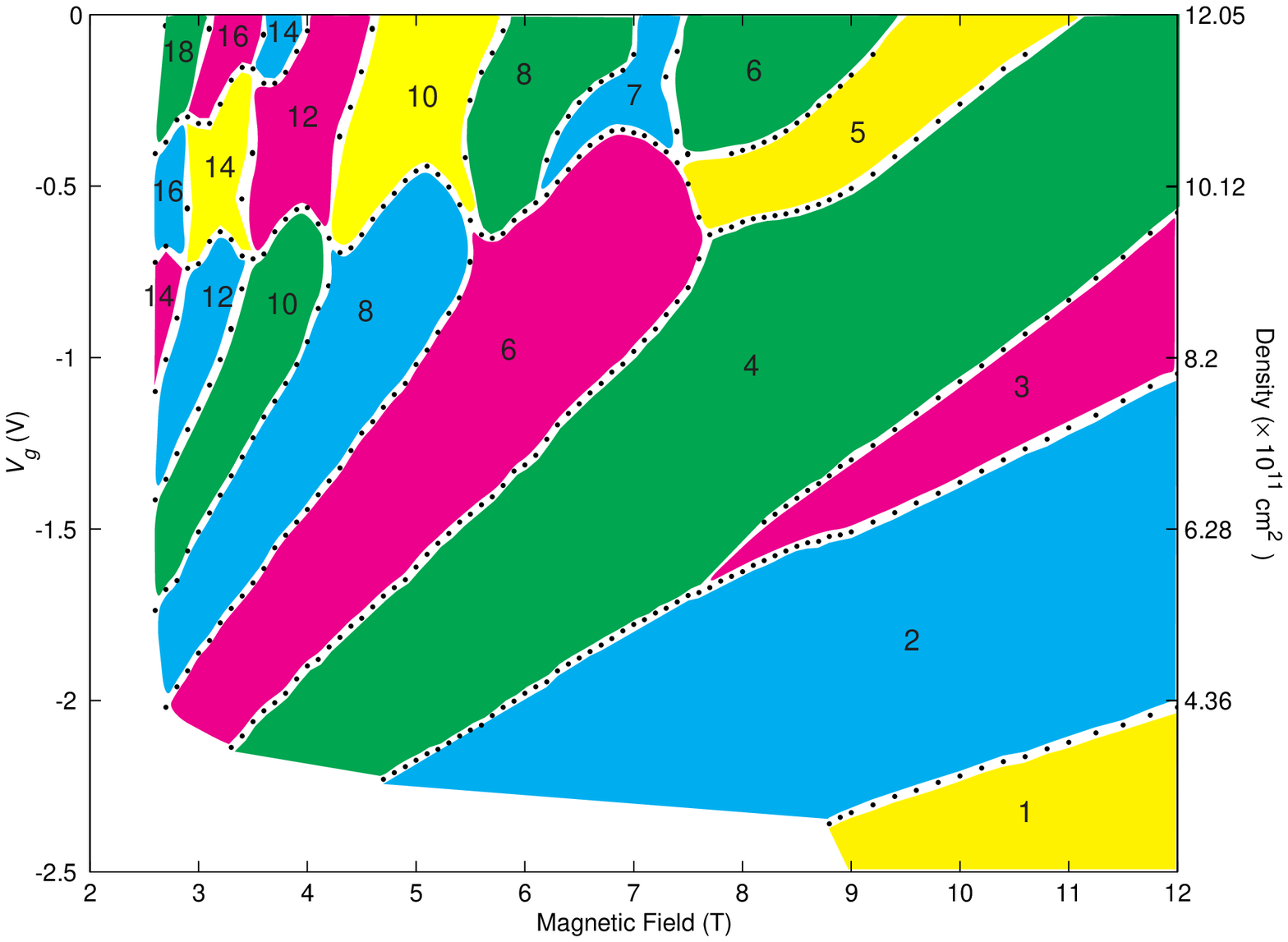, width = 9cm, clip=} 
  \caption{The experimental phase diagram in the $n$-$B$ plane.
    The data points are the phase boundaries.  $S_{xy}$ is used to label
    the various quantum Hall states.}  
  \label{exptpd} 
\end{figure}

In the low-density regime of the phase diagram (i.e., $n\le~8.5~\times
10^{11} \mbox{ cm}^{-2} $ ), with the upper subband depopulated, the
experiment phase diagram is identical to that of a one-subband system
\cite {Glozman1}. Note the transition of spin-resolved quantum Hall
states to spin-degenerate quantum Hall state at around 8~T.  This type
of level ``pinch off'' has been studied recently in detail both
theoretically \cite{Fogler,Tikofsky} and experimentally. \cite{Wong}
With the upper subband populated, the phase diagram becomes very rich
in topology. The most pronounced feature is the sawtooth-like pattern
for densities in the range between $n = 9\times10^{11} \mbox{
  cm}^{-2}$ and $11 \times10^{11} \mbox{ cm}^{-2}$. A similar pattern
can also be seen for higher densities between $n =10.5 \times10^{11}
\mbox{ cm}^{-2}$ and $12 \times10^{11} \mbox{ cm}^{-2}$. There are
also apparently ``triple'' and ``quadruple'' points which separate the
different quantum Hall phases.

Despite the complexity of this phase diagram, we find that the
``selection rules'' for the quantum Hall phase transitions are never
violated. \cite{Kivelson} According to the selection rules, the Fermi
level must cross one delocalized level at a time. For the level
counting, ``one'' is used for the spin resolved case and ``two'' is
used for the spin unresolved case.  Therefore for either a
``$V_g$-scan'' or a ``$B$-scan'', the number for $S_{xy}$ changes
either by one for crossing a spin-resolved subband level, or by two
for crossing a spin-degenerate subband level, or even conceivably by
four for crossing a spin-degenerate and subband-mixed level.

One of the striking consequences of this unusual topology is that the
differential Hall resistance (NDHR) $d\rho_{xy}/dB$ can be negative
during the $B$-scan in certain regions of the $n$-$B$ diagram. For
example a NDHR can be seen in the Fig.~1 around $B = 7$~T (in the
circled area). This NDHR is certainly unusual. In a one-band system,
only positive differential Hall resistance (i.e.~the classical Hall
resistance or in the region between two plateaus) and zero
differential Hall resistance (i.e., in the plateau region) have been
observed. For this two-band system we found, in fact, that NDHR can be
seen during a $B$-scan along a trajectory cutting through the top
portion of any sawtooth.

We present, in the left shown of Fig.~4, the evolution of the NDHR for
various $V_g$ at a fixed temperature of 70~mK for the sawtooth between
6 to 8~T. For the convenience of tracking $S_{xy}$, we have plotted
$1/R_{xy}$ (in unit of $e^{2}/h$) as the vertical axis. At $V_g =
-0.34$~V (at the tip of the tooth), a slight dip is seen at the middle
of the well developed $h/7e^2$ Hall plateau. As the $V_g$ gets more
negative, the dip shows more deviation from $h/7e^2$. The dip gets
progressively deeper and wider. At $V_g = -0.38$~V, the deviation is
the greatest while the high field portion of the $h/7e^2$ plateau is
still visible. The high-field side of this dip leads to the unusual
NDHR. As the gate voltage gets even more negative, the dip develops
into the $h/6e^2$ Hall plateau (see for example $-0.54$~V ) giving,
eventually, the normal $S_{xy}=7$ to $S_{xy}=6$ (``6-7'' in short )
quantum Hall transition.

\begin{figure}
  \epsfig{file = 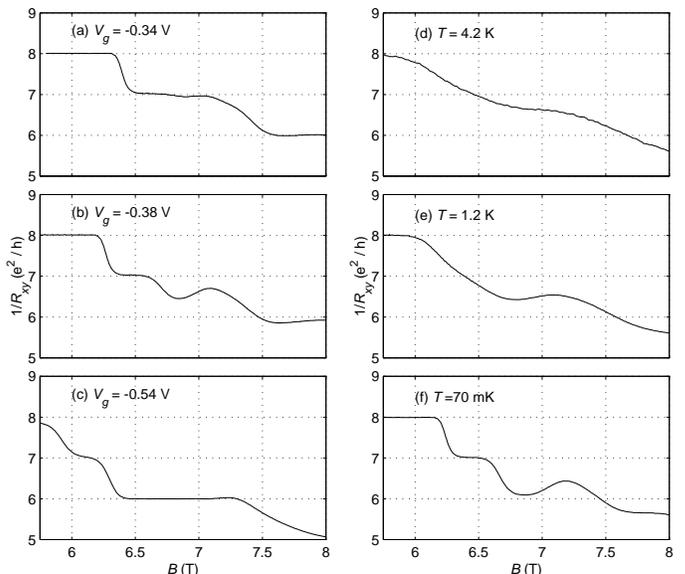, width = 9cm, clip=}
  \caption{Left side: $1/R_{xy}$ versus $B$ at three different gate
    voltages showing the evolution of the negative differential Hall
    resistance.  Right side: $1/R_{xy}$ versus $B$ at $V_g$ = -0.41~V
    at three different temperatures.}  
  \label{NDHR} 
\end{figure}

We have also investigated the temperature dependence of
$d\rho_{xy}/dB$ at $V_g = -0.41$~V. $B$-scan traces, in the right
shown of Fig.~4, were taken at $T= 4.2$~K, 1.2~K, and 70~mK. At the
highest temperature, the $h/7e^2$ plateau is not resolved and there
was no sign of the NDHR. At $T =1.2$~K, the plateau starts to form and
a small dip becomes visible near the expected positions of the
$h/7e^2$ and $h/6e^2$ plateau. As the temperature goes further down,
both $h/7e^2$ and $h/6e^2$ plateaus are well resolved and the dip
becomes deeper. The deviation has reached a value of about $h / 6.5
e^2$ at 70 mK. It is apparent that the NDHR is associated closely with
the formation of the quantum Hall states of $S_{xy}=6$ and $S_{xy}=7$
in this case.

In an attempt to understand the topological anomalies of the phase
diagram, we have performed a simple numerical calculation to account
for the effect of Landau level mixing of the two bands. In this
calculation, we have made the simple assumption that the density of
states can be modeled as two sequences of gaussian functions centered
around the Landau levels for the lower and upper subband respectively.
The width of the gaussian functions is determined from the
conductivity of the sample.\cite{Glozman2} For each and every maximum
in the density of states, the electron density is calculated at a
given magnetic field. We assume that the delocalized states lie at the
local maxima in the resulting density of states.  In this way, we can
obtain a theoretical phase diagram. Of course, in reality, the Landau
level mixing due to both the level repulsion and disorder broadening
is far more complicated than this simple assumption. Our simple
calculation is nevertheless able to produce the sawtooth-like
structure qualitatively as seen in the experimental data. Therefore,
we believe the sawtooth pattern is a result of the mixing of the
Landau levels of the two bands. Every time the two levels move towards
a crossing, the position of the delocalized states deviate from the
normal fan lines and ``float up'' in density or equivalently in
energy. At the same time, when two levels move away from the crossing,
the position of the delocalized states ``sinks down'' back to the
normal fan lines. We think this effect has the same origin as the
``floating'' observed in the one-band system in a vanishing
$B$.\cite{Glozman1,note} In our numerical calculations, the above
criterion results in a general floating up in energy of individual
delocalized states with decreasing $B$. Therefore, the unusual
sawtooth patterns are caused by delocalized states rising above their
corresponding host Landau levels. The sawtooth structure at 7.5~T can
be identified as due to the mixing of the spin-degenerate third Landau
level ($N=2$) of the first subband with the spin-up state of the
lowest Landau level of the second subband ($N^{\prime}=0\uparrow$).
The features at around 5.5~T, 4.4~T, and 3.5~T, are due to the mixing
of the ($N^{\prime}=0$) level with ($N=3, 4, 5$) levels respectively.
Another portion of the sawtooth pattern at high densities is due to
the mixing of ($N^{\prime}=1$) level with the $N=5, 6$ levels.

It is important to point out here that the sawtooth patterns indicate
a sequence of reentrant quantum Hall transitions (i.e., 7-6-7,
10-8-10, and 12-10-12 etc.). This type of reentrant quantum Hall
transitions have been proposed theoretically for single band quantum
Hall systems. \cite{Kivelson,note}  Experimentally, it has not been
seen to date. An analogous transition which has been observed is the
reentrant insulator-quantum Hall transition known as the 0-2-0
transition \cite{Jiang} or the 0-1-0 transition. \cite{Shahar} For the
present experiment, we believe the NDHR observed in certain regions
shows preliminary evidence of the long sought reentrant quantum Hall
transitions. For example, in the region between 6~T and 8~T, a
$B$-scan at the appropriate $V_g$ at the top portion of the sawtooth
is equivalent to traversing the phase diagram horizontally, thus
cutting through two sides of a sawtooth. To the left of the sawtooth,
we can identify the quantum Hall state as $S_{xy} = 7$ state. Inside
the sawtooth it is a $S_{xy} = 6$ state. To the right, within a very
narrow range of $B$, it is again a $S_{xy} = 7$ state. For the 7-6-7
transition, the Hall resistance should vary from $h/7e^{2}$ to
$h/6e^{2}$ and back to $h/7e^{2}$ with increasing $B$. However, the
$S_{xy}=6$ and the ``re-entrant'' $S_{xy}=7$ plateaus in our
experiment can not be well resolved simultaneously at a given $V_{g}$.
As a result, the values of the dip and the peak in $R_{xy}$ only
reach, at best, about $h/6.5e^{2}$ rather than $h/6e^{2}$ and
$h/7e^{2}$ respectively.  The NDHR can be considered as a precursor of
the reentrant $S_{xy} = 7$ state.  We believe one should be able to
observe the true 7-6-7 transition at lower temperatures. We however
cannot eliminate the possibility of that a true quantum Hall state
(i.e., with zero diagonal resistance and a quantized Hall plateau)
would be intrinsically prohibited by the Landau level mixing.

The authors would like to thank S.~Kivelson and D.~Orgad for helpful
discussions. This work is supported by NSF under grant \#DMR~9705439.

\end{document}